\documentclass{elsart}
\usepackage{graphicx}
\usepackage{amssymb}
\journal{}

\begin{document}

\begin{frontmatter}

\title{Optical Levitation of a Droplet under Linear Increase of Gravitational Acceleration}

\author{Masahiro I. Kohira\thanksref{chudai}},
\author{Akihiro Isomura},
\author{Nobuyuki Magome\thanksref{nbunri}},
\author{Sadaatsu Mukai\thanksref{kaiyo}},
\author{Kenichi Yoshikawa\corauthref{cor}}

\corauth[cor]{Corresponding author. FAX: +81 75 753 3779}
\ead{yoshikaw@scphys.kyoto-u.ac.jp}

\thanks[chudai]{Department of Physics, Chuo University, Kasuga,
 Bunkyo-ku, Tokyo 112-8551, Japan.}
\thanks[nbunri]{Department of Food and Nutrition,
Nagoya Bunri College, Nagoya 451-0077, Japan.}
\thanks[kaiyo]{Japan Agency for Marine-Earth Science and
Technology, Extremobiosphere Research Center, 2-15 Natsushimacho,
Yokosuka, Kanagawa 237-0061, Japan.}
\address{Department of Physics, Graduate School of Science, Kyoto University, Kyoto 606-8502, Japan\\}
\begin{abstract}
Optical levitation of a liquid droplet in gas phase was investigated
 under time-dependent change of the gravitational acceleration with
 specific flight pattern of an airplane. Through multiple trials
 under linear increase of effective gravitational acceleration,
 we performed the experiment of optical trapping of a droplet from $0.3 g_0$ to $0.9 g_0$, where $g_0 = 9.8 \ {\rm m/s^2}$. During such
 change of the effective gravitational acceleration, the trapping
 position on a droplet with the radius of ${\rm 14 \ \mu m}$ was found to be lowered by ca. ${\rm 100 \ \mu m}$. The essential
 feature of the change of the trapping position is reproduced by a
 theoretical calculation under the framework of ray optics. As far as we
 know, the present study is the first report on optical levitation under
 time-dependent gravitational change.
\end{abstract}

\end{frontmatter}

\section{Introduction}
Since the pioneering study of Ashkin on optical manipulation,
trapping by focused laser has been actively studied. It has been shown
that laser trapping serves as a useful experimental tool on the
manipulation of micro systems in liquid solutions. In contrast to
active studies on the optical trapping in liquid phase, only a few
studies on optical levitation in gas phase have been reported (see
ref.\cite{ashkin_sci_1975} and the references in ref.\cite{magome_jpc_2003}). Optical
injection can generate at most only 10 nN as theoretical limit when the
power of light is 1 W\cite{ashkin_ieee_2000}. Thus, in the case of
ground-based experiments, the size of levitation object is limited upto the
order of tens of ${\rm \mu m}$. 

As for the levitation in gas phase beside optical trapping, several methodologies have been
examined such as electrostatics\cite{rhim_rsi_1993}, electromagnetics\cite{electromagnetic}, aerodynamics\cite{aerodynamic},
and ultrasonics\cite{ultrasonic}. With these methods it is still difficult to trap
and manipulate a desired single object. 
Thus, it may be of scientific value to perform the experiment of
optical levitation in gas phase toward the optimization of the
efficiency as well as the improvement of the performance. 

We have reported optical
trapping of a water droplet with the radius of ${\rm \mu m}$ by using
highly converged laser in gas phase on the ground\cite{magome_jpc_2003}. 
In the present study, we adapt the low converged laser in order to
obtain large working distance. We will report the result of optical levitation under time
dependent change of effective gravitational acceleration with a
specific designed flight of an airplane. This study carried out in toward the future
application for the experiments under microgravity in the International Space Station (ISS).

\section{Experimental}
We performed the trapping experiments in a jet airplane (Mitsubishi
MU-300, operated by Diamond Air Service Co., Aichi, Japan). The
flight pattern was selected so as to change the acceleration of the airplane
along vertical axis $d^2H/dt^2$ linearly as shown in Fig.1.
\begin{figure}
\includegraphics[scale=0.45]{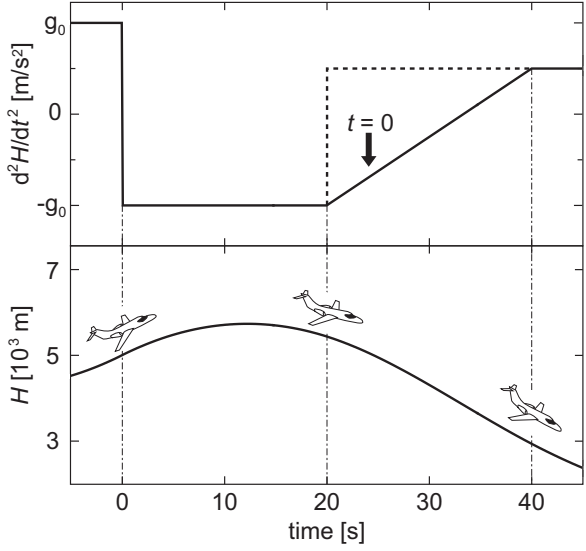}
\caption{Schematic diagram of the flight pattern. The lower diagram
 shows the time-trace of height $H$ from the ground, and the upper one
 shows acceleration $d^2H/dt^2$ along vertical axis. 
 The broken line in the upper diagram shows acceleration in the case of the conventional parabolic flight pattern. In the present study,
 because of the technical reason we analyze the region of the linear
 increase of the effective gravitatinal acceleration, where the point, $t=0$, is the time-zero adapted
 in Figs. 3 and 4.}
\end{figure}
It is to be noted that such a change of the gravitational acceleration is
largely different from the conventional parabolic flight (see the broken line in the Fig.1) usually adapted
for the experiment of ``microgravity''.

Experimental setup is schematically illustrated in Fig.2.
\begin{figure}
\includegraphics[scale=0.30]{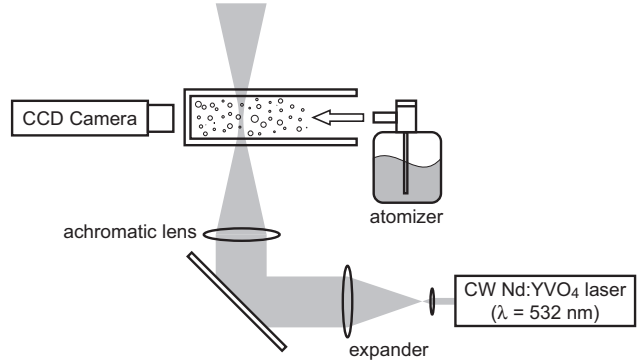}
\caption{Schematic diagram of the experimental setup.}
\end{figure}
In order to improve the dispersibility of droplets, aqueous solution
containing 30 \ vol$\%$ of ethanol was used. The droplets were injected
into glass cell (${\rm 10 \ mm \times 10 \ mm \times 50 \ mm}$) with an atomizer during
effective gravitational acceleration in the airplane $g$ ($= d^2H/dt^2 + g_0$,
where the gravitational acceleration on the ground $g_0 = 9.8 \ {\rm m/s^2}$) was
$0.01 g_0$ (e.g. between 0 and 20 s in Fig.1). For optical levitation,
slightly converged continuous wave laser beam (wavelength ${\rm \lambda =
532 \ nm}$, TEM$_{00}$ mode, Millennia, Spectra Physics) at 150 mW was focused
from the beneath into the cell with achromatic lens. The convergence angle
was 5 degrees. The time course on the manner of optical trapping of droplets
was monitored with a CCD camera. The experiments are carried out at
ca. 298 K.

\section{Results and Discussion}
Fig.3 shows the result of optical levitation on a droplet, where the time
$t=0$ and position $z=0$ are taken at the moment at which the droplet is begun to be trapped.
\begin{figure}
\includegraphics[scale=0.35]{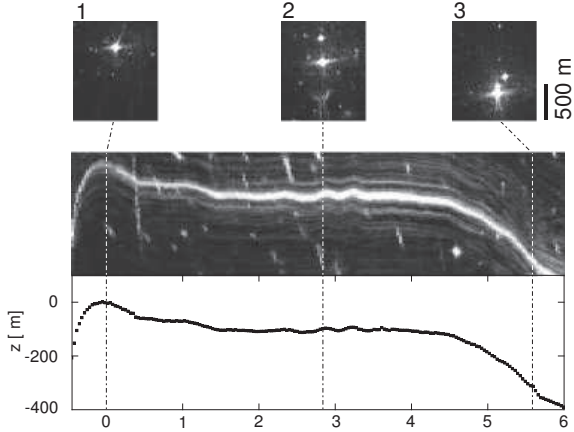}
\caption{(Top): Change of the optical microscopic images on the trapped
 droplet during the linear increase of the effective gravitational
 acceleration. (Middle): Spatio-temporal representation on the trapped
 droplet. (Bottom): Time dependent change of the height $z$ of the trapped
 droplet.}
\end{figure}
Without laser irradiation, these droplets tended to fall down and the
rate of the downward motion was increased with the increase of the
effective gravitational acceleration $g$. With laser irradiation from the
beneath, under microgravity condition, droplets on the optical corn were
scattered to the laser direction. When $g=0.3g_0$, a droplet was spontaneously trapped. As in the figure, the trapping position was gradually lowered accompanied with the time development which is correlated with the increase of $g$. The spatio-temporal image indicates the gradual change of the trapping position, where many other floating droplets outside the trapping position were continuously falling down.

Fig.4(a) shows the change of the trapping position, $z$, with the
change of the effective gravitational acceleration, where $z$ changes
gradually until $t$ = ca. 4.5 s. Then, the droplet tends to fall down,
which is attributed to the escape from the trapping potential due to the
increase of $g$. Open circles in Fig.4(b) shows the correlation between
$z$ and the relative value $g/g_0$.

\begin{figure}
\includegraphics[scale=0.45]{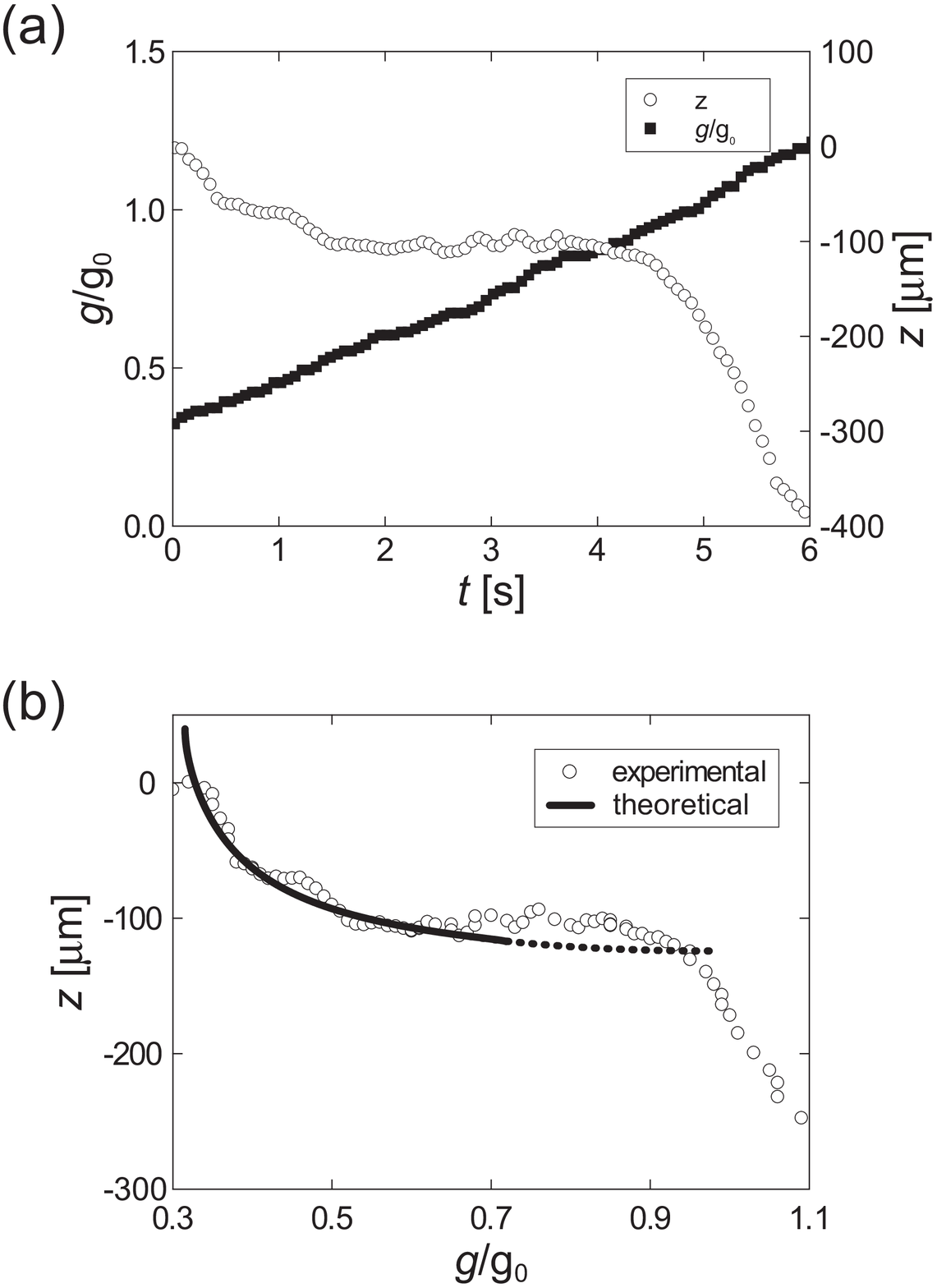}\\
\includegraphics[scale=0.70]{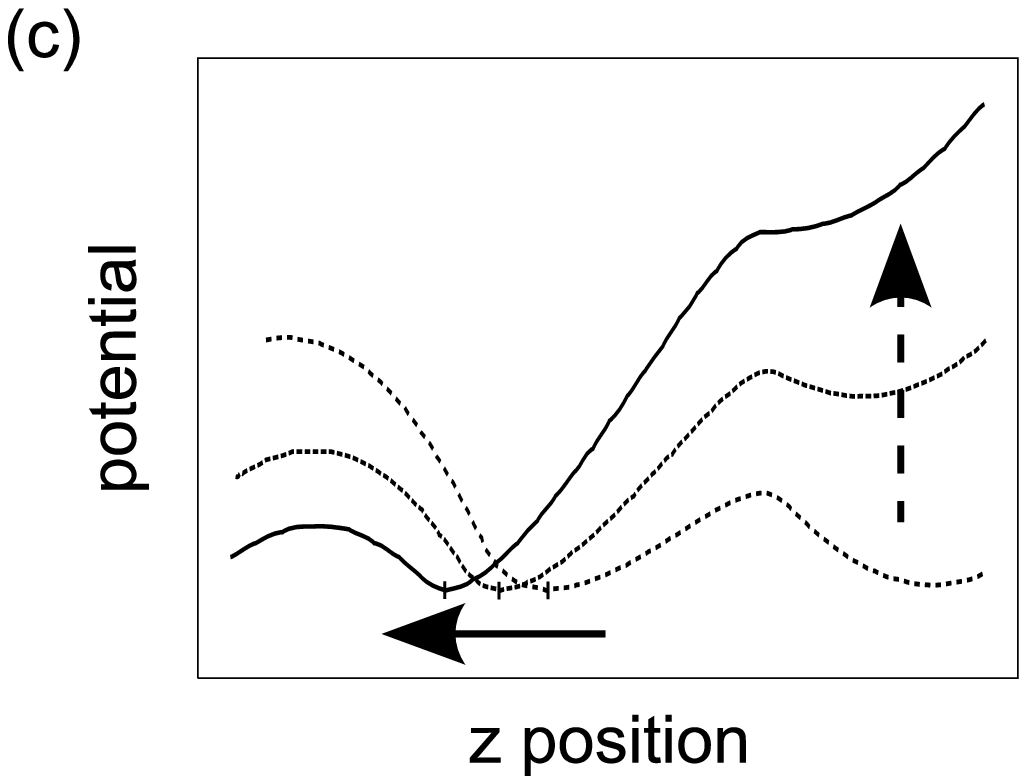}
\caption{(a) Time trace of the effective gravitational acceleration in
 the flight given in Fig.1, where $g = d^2H/dt^2 + g_0$; $g_0$ and $g$ are the
 gravity acceleration on the ground and effective gravitational
 acceleration in the airplane. The longitudinal axis is given as the
 relative value $g/g_0$. The change of the height $z$ of the trapped droplet
 is also shown. (b) Change of the height $z$ of the trapped droplet with
 respect to $g/g_0$. Experimental and theoretical results are given as open
 circles and solid line, respectively. The broken line is the expected curve
 from a theoretical consideration. In the theoretical calculation,
 following parameters are adapted: laser power $P {\rm = 150\ mW}$, convergence
 angle $\phi$ = 7.0 degrees, injected ratio of laser to the lens $\sigma/L =
 1.5$, droplet radius $r = 14.0\ {\rm \mu m}$, refractive index of a medium $n_1 =
 1.00$, refractive index of a droplet $n_2 = 1.35$, density of a droplet
 $\rho = 9.65 \times {\rm 10^2\ kg/m^3}$, gravitational acceleration on the ground $g_0
 {\rm = 9.80\ m/s^2}$. (c) Schematic illustration of the relationship
 between the droplet's motion and the shift of potential minimum. The
 broken arrow showes the change of the net potential profile including
 both the optical and gravitational contribution: The solid
 arrow shows the shift of potential minimum position. 
}
\end{figure}

In the present experiment, it was difficult to estimate the actual size
of the trapped droplet from the image itself, because of the low
magnification of the objective lens on the CCD camera. Thus, we will
have the discussion to evaluate the size of the trapped droplet from the experimental
observation on the change of the trapping position depending on the
change of the effective gravitational acceleration. The motion equation
of the droplet is:
\begin{equation}
m \ddot{z} = F_g + F_l + F_v
\end{equation}
the inertial force $m \ddot{z}$ ($m$: mass of a droplet), the gravitational force $F_g$,
the force induced by converged laser $F_l$ along the optical axis and
the viscous force $F_v$. Since the radius of the trapped droplet is on the
order of 10\ ${\rm \mu m}$, we can assume the viscous limit: $m \ddot{z}
\cong F_v$, and obtain the relation:
\begin{equation}
F_g + F_l = 0
\end{equation}
\begin{equation}
F_g = - \frac{4}{3} \pi r^3 \rho g
\end{equation}
where $g$: effective gravitational acceleration and $\rho$: density of a droplet.
Additionally, the radius $r$ of the droplet is enough larger than the wave
length $\lambda$. So we can use the ray optics regime in calculating
$F_l$,
\begin{equation}
F_l = F_l(r,z;P,\phi,\frac{\sigma}{L},n_1,n_2)
\end{equation}
where $r$: radius of a droplet, $z$: distance between focal point of laser and the center of a
droplet, $P$: laser power, $\phi$: convergence angle of laser, $\sigma /L$:
injected ratio of TEM$_{00}$ mode laser to the lens ($\sigma$ is deviation of
TEM$_{00}$ mode laser and $L$ is lens radius), 
$n_1$: refractive index of a medium, $n_2$: refractive index of a
droplet, $Q_z$: trapping efficiency along optical axis, and $c$: velocity
of light, respectively. In this case, $r$ and $z$ are variable, and the
other parameters are fixed.

$F_l$ is found as follows. In the framework of the
ray optics, a TEM$_{00}$ mode laser can be divided to rays which are
suffixed with $i$ and each power $n_1 P_i / c$ is related to $\sigma /L$. Each
rays hit the surface of the droplet at different incident angles
$\phi_i$ ($0 \leq \phi_i \leq \phi$), repeat reflections and transmissions in
the droplet until the intensity of rays reduce to zero of limit, and
give the momentum with a certain efficiency along z-axis, $Q_i=Q_i(r,z;\phi_i,n_1,n_2)$: 
\begin{eqnarray}
Q_{i} = \sin \phi_i \left\{ R_i \sin 2 \theta_i -
	n	       \frac{T_i^2[\sin(2\theta_i-2r_i)+R_i\sin2\theta_i]}{1+R_i^2+2R_i\cos2r_i}
		      \right\}  \nonumber\\
+ \cos \phi_i \left\{ 1+R_i \cos 2
		      \theta_i -
		      \frac{T_i^2[\cos(2\theta_i-2r_i)+R_i\cos2\theta_i]}{1+R_i^2+2R_i\cos2r_i} \right\} \label{eq:qi}
\end{eqnarray}
where $\theta_i$: incident angle, $r_i$: refractive angle, $R_i$:
reflection coefficient and $T_i$: transmission coefficient. These
parameters are obtained by considering geometric relation between the
droplet and the direction of a beam (see ref.\cite{ashkin_bpj_1992}).
Now, assuming the spherical droplet with radius $r$, whose center is
located at distance $z$ from the focal point of the laser, we calculate
the details of reflections and transmissions concerning all the paths of
laser beam.

The total force $F_l$ is 
\begin{equation}
F_l = \sum_{i} \frac{n_1 P_i}{c} Q_i = \frac{n_1 P}{c} Q_z 
\end{equation}
where $Q_z=Q_z(r,z; \phi,\frac{\sigma}{L},n_1,n_2)$. Thus, we obtain the equation:
\begin{equation}
\frac{4}{3} \pi r^3 \rho g = \frac{n_1 P}{c} Q_z\label{eq:net_force}
\end{equation}
By changing the radius $r$ in the above equations, we have tried to find
the optimal curve to fit the experimental observation.
Fig.4(b) shows the change of the trapping position, i.e., minimum on the
potential, together with the experimental result on the height $z$
of the trapped droplet with respect to the relative value $g/g_0$. From
this result, we can qualitatively understand the motion of the trapped
droplet as the shift of the potential minimum point (see Fig.4(c)).

From the above analysis, it has become clear that the essential feature
of the optical levitation has been theoretically reproduced in a
satisfactory manner. Using eq.(\ref{eq:net_force}) at the critical value at
$g = 1.0 g_0$ to levitate the droplet which is found by experimental
result, we obtain maximum trapping efficiency along optical axis $Q_z$ as 0.22.
It is noted that this efficiency is one order of magnitude greater than
those in previous reports (see the table 1 of
ref.\cite{magome_jpc_2003}), except to our recent result on the
ground\cite{magome_jpc_2003}.
Since the water droplet was atomized in the microgravity
condition in this experiment, the inertia of the droplet was small and reduced by viscous
drag. Additionally, the viscous limit was
accomplished by slowly changing the gravitional acceleration. We could,
thus, evaluate the critical value of the gravitational acceleration on the
optidal levitation.

In the present study we have adapted the small converged laser, indicating the
experimental condition with rather large working distance on the order
of cm. The combination of the small converged laser and
microgravity condition may afford interesting experimental system on the
optical levitation. We have a plan to improve the experimental system by
using multiple laser sources with small converged laser.

\section{Conclusion}
We investigated the change of the trapping position of a droplet under
time-dependent change of the gravitational acceleration. It was shown that theoretical
calculation based on ray optics reproduced the experimental
trend. The present result may contribute in designing the manipulating
system on the International Space Station (ISS), including the
experiments of protein crystal growth\cite{protein_crystal_growth} and water droplet
growth\cite{magome_jpc_2003}.

\section{Acknowledgements}
The authors thank Mr. S. Watanabe for helpful suggestions and
Ms. Hayata, Messrs. Fujii, Kawakatsu, and Takahashi for technical
assistance. This research was supported by the
Grant-in-Aid for the 21st Century COE ``Center for Diversity and
Universality in Physics'' from the Ministry of Education, Culture,
Sports, Science and Technology (MEXT) of Japan, and ``Ground-based
Research Announcement for Space Utilization'' promoted by the Japan
Space Forum.

\bibliographystyle{elsart-num}
\bibliography{draft_eprint}

\begin{thebibliography}{1}
\expandafter\ifx\csname url\endcsname\relax
  \def\url#1{\texttt{#1}}\fi
\expandafter\ifx\csname urlprefix\endcsname\relax\def\urlprefix{URL }\fi

\bibitem{ashkin_sci_1975}
A.~Ashkin, J.~Dziedzic, Optical levitation of liquid drops by radiation
  pressure, Science 187 (1975) 1073.

\bibitem{magome_jpc_2003}
N.~Magome, M.~I. Kohira, E.~Hayata, S.~Mukai, K.~Yoshikawa, Optical trapping of
  a growing water droplet in air, J. \ Phys. \ Chem. \ B 107 (2003) 3988.

\bibitem{ashkin_ieee_2000}
A.~Ashkin, History of optical trapping and manipulation of small-neutral
  particle, atoms, and molecules, IEEE Journal on Selected Topics in Quantum
  Electronics 6 (2000) 841.

\bibitem{rhim_rsi_1993}
W.~K. Rhim, S.~K. Chung, D.~Barber, K.~F. Man, G.~Gutt, A.~J. Rulison, R.~E.
  Spjut, Rev.\ Sci.\ Instrum. 64 (1993) 2961.

\bibitem{electromagnetic}
C.~Notthoff, H.~Franz, M.~H. D.~M. Herlach, D.~Holland-Moritz, W.~Petry,
  Electromagnetic levitation apparatus for investigations of the phase
  selection in undercooled melts by energy-dispersive x-ray diffraction, Rev.\
  Sci.\ Instrum. 71 (2000) 3791.

\bibitem{aerodynamic}
P.~F. Paradis, F.~Babin, J.~M. Gagn\'{e}, Study of the aerodynamic trap for
  containerless laser materials processing in microgravity, Rev.\ Sci.\
  Instrum. 67 (1996) 262.

\bibitem{ashkin_bpj_1992}
A.~Ashkin, Forces of a single-beam gradient laser trap on a dielectric sphere
  in the ray optics regime, Biophys.\ J. 61 (1992) 569.

\end{thebibliography}
\end{document}